\documentclass[nohyper,12pt]{JHEP3}
\usepackage{amssymb}

\def\be{\begin{equation}}
\def\ee{\end{equation}}
\newcommand{\eeq}[1]{\label{#1}\end{equation}}
\def\bea{\begin{eqnarray}}
\def\eea{\end{eqnarray}}

\def \rif  {right--invariant field}
\def\coor{coordinate}
\def\eqn{equation}
\def\tfn{transformation}

\def\cond{condition}
\def\bc{boundary condition}

\def\+{{+\!\!\!+}}
\def\-1{^{-1}}

\def\unit{{\bf 1}}

\def\el{\stackrel{\rm L}{e}}

\def\real{{\mathbb R}}

\def\dpm{{\partial_\pm}}
\def\bndry{|_{\sigma=0,\pi}}

\def\sm{$\sigma$--model}
\def\pl{Poisson--Lie }
\def\pltp{Poisson--Lie T--pluralit}
\def\pltd{Poisson--Lie T--dualit}
\def\dd{Drinfel'd double}

\def\text{}

\def\wh{\widehat}
\def\dotplus{\stackrel{{\bf.}}{+}}

\def\rd{{\rm d}}

\def\Ran{{\rm Ran}}
\def\rank{{\rm rank}}

\def\cf{{\cal {F}}}

\def\cd{{\mathfrak d}}

\def\cg{{\mathfrak g}}
\def\tcg{{\tilde {\mathfrak g}}}

\def\ttil{\tilde{T}}
\def\htil{\tilde{h}}
\def\that{\widehat{T}}
\def\ghat{\hat{g}}

\def\rpmg{\rho_\pm(g)}
\def\rpg{\rho_+(g)}
\def\rmg{\rho_-(g)}

\def\rptilh{\tilde \rho_+(\tilde{h})}
\def\rmtilh{\tilde \rho_-(\tilde{h})}

\def\ch{{\cal H}}

\def\gluingmgen{R}
\def\gluingmphi{R}

\def\gluingm{R_\rho}
\def\gluingmh{\wh {R_\rho}}
\def\gluingoper{{\cal R}}
\def\gluingoperh{\widehat {\cal R}}

\def\ld{{\mathfrak D}}

\def\PP{p}
\def\QQ{q}
\def\RR{r}
\def\SSS{s}

\def\brane{{\cal D}}
\def \ca{{A_c}}\def \cb{{B_c}}
\def\Pic{\Pi_c}
\def\dr{\partial_\rho}
\def\isospace{{V_{{\cal D}}}}

\title{Description of D--branes invariant under the Poisson--Lie T--plurality}

\author{Ladislav Hlavat\'y and Libor \v{S}nobl\thanks{This work was supported by the project No.\ 202/06/1480 of the Grant
Agency of the Czech Republic and by the research plans LC527  and MSM6840770039 of the
Ministry of Education of the Czech Republic.}\\
Faculty of Nuclear Sciences and Physical Engineering,
\\ Czech Technical University,
\\ B\v rehov\'a 7, 115 19 Prague 1, Czech Republic
\\ \email{Ladislav.Hlavaty@fjfi.cvut.cz, Libor.Snobl@fjfi.cvut.cz} 
}

\abstract{We write the conditions for open strings with charged endpoints in the language of gluing matrices.
We identify constraints imposed on the gluing matrices that are essential in this setup and investigate the
question of their invariance under the Poisson--Lie T--plurality transformations. We show that the chosen set of constraints is
equivalent to the statement that the lifts of D--branes into the Drinfel'd double are right cosets with respect to a
maximally isotropic subgroup and therefore it is invariant under the Poisson--Lie T--plurality transformations.}

\keywords{Sigma Models, Boundary Conditions, String Duality}

\begin{document}

\section{Introduction} \label{Introduction}
In our previous paper \cite{ahs:jmp08}, the \tfn{} of worldsheet boundary conditions for nonlinear sigma
models under the \pltp y \cite{klse:dna,unge:pltp} was investigated and a formula for the \tfn {} of gluing 
matrices was presented there. Boundary conditions were formulated in terms of a so--called gluing matrix that was subjected to a set of constraints originally formulated for supersymmetric models in \cite{ALZ1,ALZ2}. Abelian T--duality of such models (and also of their purely bosonic analogues) was studied in \cite{ALZ3} and later also partially extended to Poisson--Lie T--duality context in \cite{ARE}.  Unfortunately, we have shown in \cite{ahs:jmp08} that some of the constraints are not preserved under 
the \pl{} \tfn{} (even in the simplest non--Abelian T--duality context). 

In this paper we present a restricted set of constraints for the gluing matrix that does not disqualify the
interpretation of corresponding boundary condition in terms of D-branes and simultaneously preserves its
validity under the \pl \tfn s. It means that well defined D-branes formulated in this way transform into
well defined D-branes again under the \pltp y.

The existence of such description was to be expected because there exists a different, geometric formulation
of the same problem based on the geometry of D--branes lifted into the \dd{} by C.~Klim\v c\'{\i}k and P. \v
Severa in \cite{klse:pltdosdb,klse:osdbwm}. The open problem was how to express their formulation in the
language of gluing matrices, i.e., how their boundary conditions manifest themselves on the level of
original \sm s.

The paper is structured as follows. Firstly, we review and modify the formulation of boundary conditions in terms
of gluing matrices (or operators) $\gluingoper$.  Secondly, we recall some of the basic properties of \pl{} T--duality and plurality and how the gluing matrices transform. Thirdly, we demonstrate a few examples we have used in the search for consistency constraints on $\gluingoper$ preserved under \pl{} transformations. Next, we rewrite the constraints on $\gluingoper$ in an equivalent form suitable for further computations (i.e. without projectors). Finally, we lift the D--branes into the \dd{}, study how the boundary conditions manifest themselves there, demonstrate the connection with the description in \cite{klse:pltdosdb} and show the invariance of our constraints.

\section{Boundary conditions and D--branes}
\label{Boundary} We investigate the boundary conditions for \eqn s of motion of nonlinear sigma models given
by the action\footnote{We use a bit unusual notation that $\partial_\pm\phi^\mu$ form row vectors of the
derivatives of $\phi$, therefore matrices of operators in our notation may differ
 by a transposition from expressions in other papers. The dot denotes matrix multiplication,  $t$ denotes transposition, $X^{-t}\equiv
(X^t)^{-1}$.}
\begin{equation} S_{\cf}[\phi]=\int_\Sigma d^2x\,
\partial_- \phi^{\mu}\cf_{\mu\nu}(\phi)
\partial_+ \phi^\nu =\int_\Sigma d^2x\,
\partial_- \phi\cdot\cf(\phi)\cdot
\partial_+ \phi^t \label{sigm1} \end{equation}
 where $\cf$ is a tensor field on a Lie group $G$ and the functions $ \phi^\mu:\ \Sigma \subset \real^2 \rightarrow \real,\
\mu=1,2,\ldots,{\dim}\,G$ are obtained by the composition $\phi^\mu=y^\mu\circ g $ of a map
 $g:\Sigma\rightarrow G$ and components of a coordinate map $y$ of a neighborhood $U_g$ of an element $g(x_+,x_-)\in G$.
For the purpose of this paper we shall assume that the worldsheet $\Sigma$ has a topology of a strip
infinite in $\tau\equiv x_+ + x_-$ direction, $\Sigma=\real\times\langle 0, \pi \rangle$ and $x_+,x_-$ are
light--cone coordinates on $\Sigma$.

We impose the boundary conditions for open strings in the form of the gluing operator $\gluingoper$ relating
left and right derivatives of the field $g:\Sigma \rightarrow G$ on the boundary of $\Sigma$,
\begin{equation}
\label{bcing} \partial_- g \bndry = \gluingoper\, \partial_+ g \bndry,\ \ \ \ \ \sigma\equiv x_+ - x_-.
\end{equation}
We denote the matrices corresponding to the operator $\gluingoper$ on $T_gG$ in the bases of coordinate
derivatives $\partial_{y^\mu}$ as $\gluingmgen$, e.g.\footnote{Similarly we shall distinguish operators from
their matrices by the calligraphic script used. This does not apply to tensorial expressions $\cf,{\cal
G},\ch$.},
\begin{equation}
\partial_-\phi\bndry=\partial_+\phi\cdot \gluingmphi\bndry . \label{cphi}
\end{equation} The explicit form of the operator $\gluingoper$ in
principle yields the embedding of a brane in the target space which is in this case the Lie group $G$.

When varying the action (\ref{sigm1}) we shall impose vanishing of boundary terms
\begin{equation}\label{varbound}\delta \phi\cdot  ({\cal G}\cdot  \partial_\sigma \phi^t + \ch\cdot \partial_\tau\phi^t )\bndry =0,
\end{equation} where ${\cal G}$ and $\ch$ are symmetric and antisymmetric part of the tensor field $\cf$.
We shall assume that the ends of an open string move along a D--brane  -- submanifold $\brane\subset G$
-- so that both $\delta \phi\bndry\in T_g\brane$ and $\partial_\tau\phi\bndry\in T_g\brane$.
Let ${\cal N}$ be a projector $T_g G\rightarrow T_g\brane$ so that
\begin{equation}\label{dfiN}
\delta \phi\bndry={\cal N}(\delta \phi\bndry),\ \ \
\partial_\tau\phi\bndry={\cal N}(\partial_\tau\phi\bndry).
\end{equation}
From Eqs. (\ref{cphi}) and (\ref{dfiN}) we may express the defining properties of ${\cal N}$ as
$$ {\cal N}\circ(\gluingoper + id)=(\gluingoper + i d), \; {\cal N}^2={\cal N}, \; \Ran \,{\cal N} = \Ran \,(\gluingoper + id), $$
i.e.
\begin{equation}\label{Nprop}
   (\gluingmgen+\unit)\cdot N   =  \gluingmgen+\unit,\; N^2=N,\; \rank \, N=\rank \,(\gluingmgen+\unit ).\end{equation}
We should stress that these properties do not specify the projector ${\cal N}$ uniquely since its kernel is not determined.
As it will become clear later on, we may consider all such projectors equivalent for any sensible use in physics.

We can rewrite the \eqn{} (\ref{varbound}) as
\begin{equation}\delta \phi\cdot  N \cdot
({\cal F}\cdot  \partial_+ \phi^t - {\cal F}^t \cdot \partial_-\phi^t )\bndry =0, \end{equation}
which after the use of Eq.\ (\ref{cphi}) becomes
\begin{equation}
\label{bEoMorig} \delta \phi \cdot N \cdot ({\cal F}  - {\cal F}^t \cdot \gluingmphi^t ) \cdot
\partial_+ \phi^t \bndry =0.
\end{equation}
Because $\delta \phi \cdot N$ and  $\partial_+ \phi^t$ are not further restricted, we find
\begin{equation}\label{LiepiEpiR}
N \cdot ({\cal F}  - {\cal F}^t \cdot \gluingmgen^t ) =0. \end{equation} Besides that there are conditions
for $N$ and $\gluingmgen$
\begin{eqnarray}
\nonumber N_\kappa^{\,\,\,\mu} N_\lambda^{\,\,\,\nu}\partial_{[\mu} N_{\nu]}^{\,\,\,\rho}  & = & 0, \\
\label{NGconds} \gluingmgen \cdot {\cal G}\cdot  \gluingmgen^t & = & {\cal G}
\end{eqnarray}
that follow from the condition that the projectors ${\cal N}$ in all points of $G$ define integrable
distribution and that the stress tensor of the action vanishes on the boundary (see e.g. \cite{ahs:jmp08,ALZ3,stanciu}).

In our previous paper \cite{ahs:jmp08}, we have used the formulation first presented in \cite{ALZ2}, i.e. we
have defined D--branes by virtue of Dirichlet projector ${\cal Q}$ that projects tangent vectors in a point
of $G$ onto the space normal to the D--brane going through this point and  the normal space was identified
with the eigenspace of $\gluingoper$ with the eigenvalue $-1$, i.e.,
\begin{equation}
\label{Qprop} Q^2=Q,\  Q \cdot \gluingmgen = -Q.\end{equation} The Neumann projector ${\cal N}$, which
projects onto the tangent space of the brane was then defined as complementary to ${\cal Q}$, i.e.
$$N:=\unit-Q.$$ The Eq. (\ref{Qprop}) is then equivalent to
$$ N^2=N, \ N\cdot(R+\unit)=R+\unit. $$
In order to get an agreement with Eq. (\ref{Nprop}) we had to assume that the geometrical and algebraic
multiplicities of the eigenvalue $-1$ are equal. This gave another condition that relates $\gluingmgen$ and
$Q$
\begin{equation}\label{multip}
Q \cdot \gluingmgen = \gluingmgen\cdot Q\end{equation} so that we got the following set of conditions
(equivalent to those in \cite{ALZ2})
\begin{eqnarray}
\label{allcond1} Q^2=Q, \ Q \cdot \gluingmgen = -Q, \ {\rm rank}\, Q & = &  \dim \ker\, (R+\unit)\\
\label{multip1} Q \cdot \gluingmgen &=& \gluingmgen\cdot Q, \\
\label{allcond2}N_\kappa^{\,\,\,\mu} N_\lambda^{\,\,\,\nu}\partial_{[\mu} N_{\nu]}^{\,\,\,\rho}  & = & 0, \\
\label{allcond3} \gluingmgen \cdot {\cal G}\cdot  \gluingmgen^t & = & {\cal G},  \\
\label{allcond4} N \cdot ({\cal F}  - {\cal F}^t \cdot \gluingmgen^t ) & = & 0.
\end{eqnarray}
We found in our previous work that the constraints for a consistent gluing operator $\gluingoper$
derived above are not in general preserved under the \pl \tfn s (see Sec. 5.2, case (100) in
\cite{ahs:jmp08}).

The situation improved a bit when we admitted that the endpoints of the string are electrically charged so that
the action must be modified by boundary terms. Such an extension in the context of Poisson--Lie T--duality of 
open strings  was already introduced in \cite{klse:pltdosdb}, in the gluing matrix language was firstly mentioned in \cite{ALZ2}. 
We have 
\begin{equation}\label{charged_action}
S_{\cf}[\phi] \rightarrow S_{\cf}[\phi] + S_{boundary}[\phi]
\end{equation}
where
\begin{equation}\label{charged_action_boundary}
S_{boundary}[\phi] = q_0 \int_{\sigma=0} A_\mu \frac{\partial \phi^\mu}{\partial \tau} {\rm d}\tau -q_0
\int_{\sigma=\pi} A_\mu \frac{\partial \phi^\mu}{\partial \tau} {\rm d}\tau
\end{equation}
corresponds to electrical charges $q_0,-q_0$ associated with the two endpoints of the string interacting
with electric field(s) present on the respective D--branes. The condition (\ref{LiepiEpiR}) is then modified
to the form \cite{ahs:jmp08}
\begin{equation}\label{allcond5}
 N \cdot \left( ({\cal F}+\Delta ) - ({\cal F}+\Delta )^t \cdot \gluingmgen^t \right)  =  0,
\end{equation}
where in local coordinates adapted to the brane\footnote{i.e.,  $\frac{\partial}{\partial {y^\mu}},\
\mu=1,\ldots,\dim$(brane) are tangential to the brane and the remaining vectors $\frac{\partial}{\partial
{y^\kappa}},\ \kappa>\dim$(brane) are transversal.} we have
\begin{equation}\label{defdelta}
 \Delta_{\mu\nu}= \frac{1}{2} \left( \frac{\partial A_\nu}{\partial y^\mu} - \frac{\partial A_\mu}{\partial y^\nu}
 \right),
\end{equation} $\mu,\nu\leq \dim$(brane) (the remaining components of $\Delta$ do not contribute to the Eq. (\ref{allcond5})). For computational simplicity we assume that $\Delta$
can be smoothly extended into a neighborhood of the brane. Because the values of $\Delta$ are physically
relevant only along the D--brane we may impose a supplementary restriction on $\Delta$ that fixes its
extension into the transversal directions
\begin{equation}\label{NDN1}
    \Delta=N\cdot\Delta\cdot N^t.
\end{equation}
The exactness of $\Delta$ along the brane (\ref{defdelta}) is locally equivalent to its closeness written in
arbitrary coordinates as \begin{equation}\label{NDN} {N_\kappa}^\nu {N_\lambda}^\rho {N_\mu}^\sigma
\partial_{[\nu} \Delta_{\rho \sigma]}
    =0.
\end{equation} Unfortunately, neither this generalized formulation of D-branes defined by the
gluing operator and interaction with the charges is preserved under the \pltp y or \pl T--duality in the
sense that there are cases when the set of conditions (\ref{allcond1})--(\ref{allcond3}),(\ref{allcond5}) and
(\ref{NDN}) holds for a \sm{} with boundary conditions given by $\gluingoper$ but not for a model
and boundary conditions obtained by the \pl \tfn{} (See Sec. 5.2, case (101) in \cite{ahs:jmp08}).

This problem forces us to reconsider the necessity of conditions (\ref{allcond1})--(\ref{allcond3}). Namely,
motivated by explicit examples in \cite{ahs:jmp08} we revisit the condition (\ref{multip1}). If this
condition holds (as is always the case when ${\cal G}$ is positive/negative definite but not in general)
then there is a canonical choice of the projector ${\cal N}$, namely, such that ${\cal N}$ is an orthogonal
projector with respect to the metric ${\cal G}$. On the other hand, if the condition (\ref{multip1}) does not
hold, one cannot choose the projector ${\cal N}$ uniquely and also it is not possible to find the so--called
adapted coordinates \cite{ALZ2}, i.e. the boundary conditions cannot be split into Dirichlet and
(generalized) Neumann directions. Although such boundary conditions may appear strange, we don't see any
reason why they should be a priori excluded from consideration.

Moreover, we shall prove that if we relax the condition (\ref{multip1}) and reformulate the other ones in such a way
that the \sm{} with boundary conditions is given by $(\cf,\gluingoper,\Delta)$ satisfying
\begin{eqnarray}
\label{allcon3} \gluingmgen \cdot {\cal G}\cdot  \gluingmgen^t & = & {\cal G},  \\
\label{allcon1} (\gluingmgen +\unit)\cdot N=(\gluingmgen +\unit), \ \ N^2&=& N, \ \ \rank\, N=\rank\, (\gluingmgen +\unit)  \\
\label{allcon2} N_\kappa^{\,\,\,\mu} N_\lambda^{\,\,\,\nu}\partial_{[\mu} N_{\nu]}^{\,\,\,\rho}  & = & 0, \\
\label{allcon4}  N \cdot \left( ({\cal F}+\Delta ) - ({\cal F}+\Delta )^t \cdot \gluingmgen^t \right)  &=&
0,\\
\label{allcon5}
    N\cdot\Delta\cdot N^t&=& \Delta,\\
\label{allcon6}  {N_\kappa}^\nu {N_\lambda}^\rho {N_\mu}^\sigma \partial_{[\nu} \Delta_{\rho \sigma]}
    &=& 0.
\end{eqnarray}
then these conditions are preserved by the \pl \tfn.

\section{Elements of \pltp y and transformation of boundary \cond s} \label{Elements}
The \pltp y was described in many papers (e.g. \cite{klse:dna,unge:pltp,tvu:pltdpi}) and we sketch here only its main features, mainly to
set the notation. The tensor field $\cf$ on the Lie group $G$ can be written
 as \begin{equation} \cf_{\mu\nu}={e_\mu}^a (g)F_{ab}(g)
{e_\nu}^b(g)\label{metorze} \end{equation} where the vielbeins ${e_\mu}^a(g)$ are components of the right-invariant Maurer--Cartan
forms ${\rm d}g g^{-1} $ and $F_{ab}(g)$ are matrix elements of bilinear nondegenerate form $F(g)$ on $\cg$, the Lie
algebra of $G$. The action of the \sm{} then reads
\begin{equation} \label{SFg}
S_{F,A}[g]=\int_\Sigma d^2x\, \rmg \cdot F(g) \cdot \rpg^t +\int_{\sigma=0}A-\int_{\sigma=\pi}A, \end{equation}
where the right--invariant vector fields $\rpmg$ are given by
\begin{equation}\label{rpm} \rpmg^a\equiv(\dpm g g\-1)^a=\dpm \phi^\mu\, {e_\mu}^a(g),\ \ \ (\dpm g g\-1)=
\rpmg \cdot T= \partial_{\pm} \phi \cdot e(g) \cdot T, \end{equation} $T_a$ are basis elements of the Lie algebra  $\cg$ and $A$ is the 1--form introduced in (\ref{charged_action_boundary}).

Similarly, the boundary conditions (\ref{bcing}) may be expressed in terms of the right--invariant fields, as
\begin{equation}
\label{origbc} \rmg\bndry = \rpg\cdot\gluingm\bndry,
\end{equation}where
\begin{equation} \gluingm =
e\-1(g)\cdot  \gluingmphi\cdot e(g).\end{equation}

The \sm s that are transformable under \pltd y can be formulated on a \dd{} $D\equiv(G|\tilde G)$, a Lie group whose
Lie algebra $\cd$ admits a decomposition $\cd=\cg\dotplus\tcg$ into a pair of subalgebras maximally isotropic with
respect to a symmetric ad-invariant nondegenerate bilinear form $\langle\, .\,,.\,\rangle $. The matrices $F_{ab}(g)$
for the dualizable \sm s are of the form
\begin{equation} F(g)=(E_0^{-1}+\Pi(g))^{-1}, \ \ \ \Pi(g)=b(g) \cdot a(g)^{-1} =
-\Pi(g)^t,\label{Fg}\end{equation}
 where $E_0$ is a constant matrix, $\Pi$ defines the Poisson structure on the group $G$, and $a(g),b(g)$ are submatrices
of the adjoint representation of $G$ on $\cd$
\begin{equation}\label{adgt} g T  g\-1\equiv Ad(g)\triangleright T=a\-1(g) \cdot T,\ \ \ \ g\tilde  T g\-1\equiv
Ad(g)\triangleright \tilde T =b^t(g) \cdot T+ a^t(g) \cdot \tilde T, \end{equation} where $\ttil^a$ are elements of
dual basis in the dual algebra $\tcg$, i.e., $\langle\,T_a ,\,\ttil^b\,\rangle=\delta_a^b $.

The bulk \eqn s of motion of the dualizable \sm s can be written as Bianchi identities for the $\tcg$--valued fields
$$ (\rho_{+})_{a} = - \rpg^bF(g)_{cb}(a(g)\-1)^c_a, \qquad (\rho_{-})_{a} = \rmg^b F(g)_{bc}(a(g)\-1)^c_a.$$
These fields can be consequently integrated in terms of suitable $\tilde h: \Sigma \rightarrow \tilde{G}$,
\begin{eqnarray}\label{emrpmtilh}\nonumber
  \rptilh_a=(\partial_+ \htil\,\htil\-1)_a &=& - \rpg^bF(g)_{cb}(a(g)\-1)^c_a,\\
   \rptilh_a=(\partial_- \htil\,\htil\-1)_a &=& \rmg^b F(g)_{bc}(a(g)\-1)^c_a.
\end{eqnarray}
This procedure defines the lift $l: \Sigma\rightarrow D$ of the solution $g: \Sigma\rightarrow G$ into the \dd{}. As a consequence, the lift satisfies
\cite{klse:dna},
\begin{equation}\label{ddeqm}
    \langle\, \dpm l  l\-1\,,{\cal E^\pm}\,\rangle =0,
\end{equation}
where $l=g \htil$ and ${\cal E^\pm}$ are two orthogonal subspaces in $\cd$, spanned by $T+E_0\cdot \tilde T$, $T-E_0^t\cdot \tilde T$, respectively. On the other hand, starting from a solution $l$ in the \dd{} we find a  corresponding solution $g$ by constructing the decomposition  $l=g \htil$.

In general, there are several decompositions (Manin triples) of a \dd{} that enable to transform one \sm{} and its
solutions into others. Let $\hat\cg\dotplus\bar\cg$ be another decomposition of the Lie algebra $\cd$. The pairs of
dual bases of $\cg,\tcg$ and $\hat\cg,\bar\cg$ are related by the linear \tfn
\begin{equation}\label{pqrs}
    \left(\matrix{T \cr\tilde T\cr} \right)= \left(\matrix{\PP&\QQ \cr \RR&\SSS \cr} \right) \left(\matrix{\that\cr
\bar T\cr} \right),
\end{equation}
where the duality of both bases requires
\begin{equation}
\left(\matrix{\PP&\QQ \cr \RR&\SSS \cr} \right)\-1=\left(\matrix{\SSS^t&\QQ^t \cr \RR^t&\PP^t \cr}
\right),\end{equation} i.e., \begin{equation} \label{dualbase}\begin{array}{ccr}
\PP \cdot \SSS^t+\QQ \cdot \RR^t&=&\unit,\\
\PP \cdot \QQ^t+\QQ \cdot \PP^t&=&0, \\
\RR \cdot \SSS^t+\SSS \cdot \RR^t&=&0.
\end{array}
\label{KQUSreln2}\end{equation}
 The \sm{} obtained by the plurality \tfn{} is then defined analogously to the original one, namely by substituting
\begin{equation} \wh F(\ghat)=(\wh E_0^{-1}+\wh\Pi(\ghat))^{-1}, \ \ \ \wh\Pi(\ghat)=\wh b(\ghat)
\cdot \wh a(\ghat)^{-1} = -\wh\Pi(\ghat)^t,\label{Fghat}\end{equation}\begin{equation} \label{E0hat} \widehat
E_0=(\PP+E_0 \cdot \RR)\-1 \cdot (\QQ+E_0 \cdot \SSS)=(\SSS^t \cdot E_0-\QQ^t) \cdot (\PP^t-\RR^t \cdot
E_0)\-1\end{equation} into (\ref{metorze}), (\ref{SFg}). Solutions of the two \sm s are related by two possible
decompositions of $l\in D$, namely
\begin{equation}\label{lgh}
    l=g \htil=\ghat \bar h.
\end{equation}
For $ \PP=\SSS=0,\ \QQ = \RR=\unit $ we get the so--called \pltd y where $ \hat G=\tilde G,\ G'=G,\ \widehat
E_0=E_0\-1$. If $G$ is non--Abelian and $ \tilde G$ is Abelian we call it non--Abelian T--duality.

The corresponding transformation of the gluing matrix $\gluingm$ under the \pltp y was found in \cite{ahs:jmp08} in
the form \begin{equation} \gluingmh={\wh F}^t(\hat g)\cdot M_-\-1\cdot F^{-t}(g)\cdot\gluingm(g)\cdot F(g)\cdot
M_+\cdot\wh F\-1(\hat g), \label{gluingmh}\end{equation}  where
\begin{equation}\label{Mpm}
 M_+ \equiv \SSS+{E_0}^{-1}\cdot \QQ,\ \ \ M_-\equiv \SSS-{E_0}^{-t}\cdot \QQ.
\end{equation}

An obvious drawback of the formula (\ref{gluingmh}) is that the transformed gluing matrix $\gluingmh$ may depend not
only on $\ghat$ but also on $g$, i.e., after performing the lift into the double $g \htil=\ghat \bar h$ it may depend
on the new dual group elements $\bar h\in \bar G$, which contradicts any reasonable geometric interpretation of the
transformed boundary conditions. A solution of this problem is that we admit gluing matrices only in the form
\begin{equation}\label{constH}
    \gluingm(g)=F^t(g)\cdot C \cdot F\-1(g),
\end{equation}
where $C$ is a constant matrix\footnote{In general, one can admit $C$ dependent on some combinations of coordinates of
$G$ that transform by \pltp y to \coor s on $\wh G$ (see \cite{ahs:jmp08}).}.
 Then $ \gluingmh$ depends only on $\ghat$.

The condition (\ref{allcon3}) requiring that $\gluingm$ of the form (\ref{constH}) preserves the metric then restricts
the form of the matrix $ C$ by
\begin{equation}\label{cece}
    C\cdot(E_0^{-1}+E_0^{-t})\cdot C^t=(E_0^{-1}+E_0^{-t}).
\end{equation}
It is an easy exercise \cite{ahs:jmp08} to show that Eq. (\ref{cece}) is preserved under the \pl{} transformation (\ref{gluingmh}).

\section{Examples of three--dimensional \sm s}\label{3dm}
{The \cond s (\ref{allcon3})--(\ref{allcon6}) can be used} in the following way. Let us  assume that the
tensor ${\cal F}$ is given. For the given metric ${\cal G}$, i.e. symmetric part of ${\cal F}$, we find
admissible gluing operators $\gluingoper$ from Eq. (\ref{allcon3}), i.e. operators orthogonal with respect
to ${\cal G}$. Then the projector $N$ is determined from Eqs. (\ref{allcon1}) and the \cond{} of
integrability (\ref{allcon2}) is checked. Finally, the 2--form $\Delta$ is obtained from
(\ref{allcon4}),(\ref{allcon5}) and we check the \cond{} (\ref{allcon6}), namely, that it is closed on the
brane. { The same procedure is then repeated for the dual or plural model with ${\wh F}$ and $\gluingmh$
given by (\ref{Fghat}) and (\ref{gluingmh})}.

As an example we shall investigate the \pl{} \tfn s of the \sm s formulated on the \dd s $D\equiv(G|\tilde
G)$, where $G$ is the Lie group corresponding to one of the nine three--dimensional Lie algebras $Bianchi$~1
- $Bianchi$~9 (for notation see e.g. \cite{snohla:ddoubles}) and $ \tilde G$ is the Abelian Lie group
corresponding to $Bianchi$~1. We shall denote these \dd s $(X|1)$ where $X$ is the number of the Bianchi
algebra.

The matrix $\Pi$ vanishes for Abelian $\tilde G$ so that $F(g)=E_0$ and
\begin{equation} \cf_{\mu\nu}={e_\mu}^a (g)(E_0)_{ab}\,
{e_\nu}^b(g).\label{metorzex1} \end{equation} We choose the constant matrix $E_0$ as
\begin{equation}\label{E0}
   E_0 =\left(\matrix{ 0  & 0&  1
   \cr 0& 1  & 0
   \cr 1   &0  & 0 \cr } \right)
\end{equation} so that we work with an indefinite metric on $G$.

Our task is to choose gluing operators $\gluingoper$ producing $\Delta$ and $N$ that satisfy the \cond s
(\ref{allcon3})--(\ref{allcon6}) and check whether the transformed gluing operators $\wh \gluingoper$ which are
expressed in the non--coordinate  frame of the \rif s by (\ref{gluingmh}), produce $\wh\Delta$ and $\wh N$ satisfying
the \cond s (\ref{allcon3})--(\ref{allcon6}) if and only if the original ones do.

The generic solution of Eq. (\ref{allcon3})  for $E_0$ given by (\ref{E0}) is
\begin{equation} \gluingm=
\left(
\begin{array}{lll}
 \beta  & \gamma  & -\frac{\gamma ^2}{2 \beta } \\
 \frac{(\alpha -\epsilon) \beta }{\gamma } & \alpha  & -\frac{(\alpha +\epsilon) \gamma }{2 \beta } \\
 -\frac{(\alpha -\epsilon)^2 \beta }{2 \gamma ^2} & \frac{1-\alpha ^2}{2 \gamma } & \frac{(\alpha
   +\epsilon)^2}{4 \beta }
\end{array}
\right), \label{C2}
\end{equation} where $\epsilon =\pm 1,$ and  $\alpha, \beta, \gamma$ are real constants such that $ \beta, \gamma\neq 0$ .

Note that the \cond s (\ref{allcon1}), (\ref{allcon4}), (\ref{allcon5}) can be calculated even in the non--coordinate
frame where $\cf=E_0$,  therefore $N_\rho$ and $\Delta_\rho$ are independent of $G$. Moreover, the \cond{}
(\ref{allcon2}) holds for all ranks of ${\cal N}$ but two and the condition (\ref{allcon6}) holds for all ranks of
${\cal N}$ but three on dimensional grounds.

Solving Eq. (\ref{allcon1}) for the above given matrix $\gluingm$ and $\epsilon =1$  we get the identity projector
${\cal N}=id$, and for $\epsilon =-1$ we get $N=e(g)\cdot N_\rho\cdot e(g)\-1$ where \footnote{This holds for generic
values of $\alpha, \beta, \gamma$. Cases $\epsilon=1,\ \alpha= - 1-2\beta$ and $\epsilon=-1,\ \alpha= 1-2\beta\pm
4\sqrt{-\beta},\ \alpha=-1 $ when forms of ${\cal N}$ are different were analyzed separately and the invariance under
T--duality was also confirmed.}
\begin{equation}\label{N2}N_\rho =
\left(
\begin{array}{ccc}
 \frac{\text{n_1} \beta ^2}{\alpha  \gamma +\gamma }+1 & \frac{\text{n_2}
   \beta ^2}{\alpha  \gamma +\gamma } & -\frac{\beta ^2 \left(\text{n_2}
   \beta  (\alpha -2 \gamma -1)+2 \left(\text{n_1} \beta ^2+\alpha  \gamma
   +\gamma \right)\right)}{2 (\alpha +1)^2 \gamma ^2} \\
 \frac{\text{n_1} \beta  (\alpha -2 \gamma -1)}{2 (\alpha +1) \gamma } &
   \frac{\text{n_2} \beta  (\alpha -2 \gamma -1)}{2 (\alpha +1) \gamma }+1 &
   -\frac{\beta  (\alpha -2 \gamma -1) \left(\text{n_2} \beta  (\alpha -2
   \gamma -1)+2 \left(\text{n_1} \beta ^2+\alpha  \gamma +\gamma
   \right)\right)}{4 (\alpha +1)^2 \gamma ^2} \\
 \text{n_1} & \text{n_2} & \frac{\beta  (-\alpha  \text{n_2}+2 \gamma
   \text{n_2}+\text{n_2}-2 \text{n_1} \beta )}{2 (\alpha +1) \gamma }
\end{array}
\right)\end{equation} and $n_1,n_2$ are arbitrary constants. The rank of the latter projector is 2.

For $\epsilon =1 $,  the \cond{} (\ref{allcon2}) is satisfied trivially  as the distribution of tangent vector spaces
of the space filling D--branes is identical with the tangent spaces of the manifold. The \cond s (\ref{allcon4}),
(\ref{allcon5}) yield
\begin{equation}\label{Deltarho2}
 \Delta_\rho=   \left(
\begin{array}{ccc}
 0 & -\frac{2 \gamma }{\alpha +2 \beta +1} & \frac{\alpha -2 \beta
   +1}{\alpha +2 \beta +1} \\
 \frac{2 \gamma }{\alpha +2 \beta +1} & 0 & -\frac{2 (\alpha -1) \beta
   }{\gamma  (\alpha +2 \beta +1)} \\
 -\frac{\alpha -2 \beta +1}{\alpha +2 \beta +1} & \frac{2 (\alpha -1)
   \beta }{\gamma  (\alpha +2 \beta +1)} & 0
\end{array}
\right),\ \  \Delta=e(g)\cdot  \Delta_\rho \cdot e(g)^t.
\end{equation}The form of $e(g)$ and therefore also the \cond{} (\ref{allcon6}) depend on $G$.

The results for $\epsilon =1 $ are:
\begin{itemize}
    \item For $Bianchi$ 1,2,$6_0,7_0$ the \cond{} (\ref{allcon6}) is satisfied for
        any gluing matrix of the form (\ref{C2}).
    \item For $Bianchi$ 3,4,5,$6_a,7_a$ the \cond{} (\ref{allcon6}) is satisfied if and only if $\alpha=1$.
\end{itemize}

If $\epsilon =-1 $,  the results are:
\begin{itemize}
    \item For $Bianchi$ 1,5 the \cond{} (\ref{allcon2}) is satisfied for
        any gluing matrix of the form (\ref{C2}).
    \item For $Bianchi$ $3,6_a$ the \cond{} (\ref{allcon2}) is satisfied if and only if
    $\beta=-1,\ \gamma=\pm2$ or $\alpha=\frac{\gamma+2\gamma\beta\,\pm\,2\beta}{\gamma\mp 2\beta}$.
    \item For $Bianchi$ $6_0$ the \cond{} (\ref{allcon2}) is satisfied if and only if
    $\alpha=1+2\beta\pm2\gamma$.
    \item For $Bianchi$ $2,4,7_0,7_a$ the \cond{} (\ref{allcon2}) is never satisfied.
\end{itemize}

It is too complicated to check the \cond s (\ref{allcon2}) and (\ref{allcon6}) for the simple groups that
correspond to $Bianchi$ $8,9$ and the generic solution of Eq. (\ref{cece}). Nevertheless, we can calculate them at least
for a particular gluing matrix
\begin{equation} \gluingm=\left(\matrix{0&0&\frac{1}{\beta}  \cr 0 & 1& -
  \,
\frac{\alpha }{\beta } \cr \beta  & \alpha  & - \frac{{\alpha }^2}{2\, \beta } \cr  }  \right) \label{C1}
\end{equation}
that is a special solution of Eq. (\ref{cece}). Solving Eq. (\ref{allcon1}) for the above given matrix $\gluingm$ we
get the projector \begin{equation}\label{N1}  N_\rho= \left(
\begin{array}{lll}
 \frac{n}{\beta }+1 & 0 & \frac{n+\beta }{\beta ^2} \\
 -\frac{n \alpha }{2 \beta } & 1 & -\frac{\alpha  (n+\beta )}{2 \beta ^2}
   \\
 -n & 0 & -\frac{n}{\beta }
\end{array}
\right),
\end{equation}where $n$ is an arbitrary constant. Rank of this projector is 2
so that the \cond{} (\ref{allcon6}) is satisfied trivially and
\begin{itemize}
    \item  For $Bianchi$ $8$ the \cond{} (\ref{allcon2}) is satisfied if and only if
    $\alpha=\pm 2 \sqrt{\beta^2-1}$.
    \item For $Bianchi$ $9$ the \cond{} (\ref{allcon2}) is never satisfied.
\end{itemize}

\subsection{Non--Abelian T--duality} As a next step,  we shall investigate the constraints for  the dual gluing
matrices obtained by the \pltd y that interchanges $G$ and $\tilde G$. We have proven in \cite{ahs:jmp08} that the
so--called conformal \cond{} (\ref{allcon3}) is preserved by the \tfn{} (\ref{gluingmh}) so it is not necessary to
check it. For the models on the \dd s $(X|1)$, the \pltd y reduces to the non--Abelian T--duality and the gluing
matrices of the dual models are
\begin{equation}
    \gluingmh=-{\wh F}^t(\hat g)\cdot E_0^t\cdot  C \cdot E_0\-1 \cdot\wh F\-1(\hat g)=
   - \left(\unit-E_0^{-t}\cdot\hat\Pi(\hat g)\right)\-1\cdot C \cdot
    \left(\unit+E_0^{-1}\cdot\hat\Pi(\hat g)\right). \label{gluingmhtdual}
\end{equation}
They  depend on the choice of $G$ which determines the matrices $\wh\Pi$. The projectors
$\wh{\cal N}$ are obtained from (\ref{allcon1}) and it turns out that the rank of the projector $\wh{\cal N}$ is
independent of $G$. For $\epsilon=1$ it is equal to 2 while for $\epsilon=-1$ it is equal to 3. It means that for
$\epsilon=1$ the nontrivial condition is (\ref{allcon2}) while for $\epsilon=-1$ it is the \cond{} (\ref{allcon6}).

 For the matrix (\ref{C2}) and  $\epsilon=1$  we get:
\begin{itemize}
    \item $Bianchi$ 1,2,$6_0,7_0$: The \cond{} (\ref{allcon2}) for $\gluingoperh$ is satisfied for
        any gluing matrix of the form (\ref{C2}).
    \item $Bianchi$ 3,4,5,$6_a,7_a$ : The \cond{} (\ref{allcon2}) for $\gluingoperh$ is satisfied if and only if $\alpha=1$.
\end{itemize}

For the matrix (\ref{C2}) and   $\epsilon=-1$ we get:
\begin{itemize}
    \item $Bianchi$ 1,5: The \cond{} (\ref{allcon6}) for $\gluingoperh$ is satisfied for
        any gluing matrix of the form (\ref{C2}).
    \item $Bianchi$ $3,6_a$ : The \cond{} (\ref{allcon6}) for $\gluingoperh$ is satisfied if and only if
    $\beta=-1,\ \gamma=\pm2$ or $\alpha=\frac{\gamma+2\gamma\beta\,\pm\,2\beta}{\gamma \mp 2\beta}$.
    \item $Bianchi$ $6_0$ : The \cond{} (\ref{allcon6}) for $\gluingoperh$  is satisfied if and only if
    $\alpha=1+2\beta\,\pm2\,\gamma$.
    \item $Bianchi$ $2,4,7_0,7_a$ : The \cond{} (\ref{allcon6}) for $\gluingoperh$  is never satisfied.

\end{itemize}

For the matrix (\ref{C1}) the projectors $\wh{\cal N}$ obtained from (\ref{allcon1}) have the rank equal to 3
so that the \cond{} (\ref{allcon2}) is satisfied trivially  and for:
\begin{itemize}
    \item $Bianchi$ $8$ the \cond{} (\ref{allcon6}) is satisfied if and only if
    $\alpha=\pm 2 \sqrt{\beta^2-1}$.
    \item $Bianchi$ $9$ the \cond{} (\ref{allcon6}) is never satisfied.
\end{itemize}

Comparing the above given results with those in the previous subsection we see that the \cond s
(\ref{allcon3})--(\ref{allcon6}) are preserved under the non--Abelian T--duality. We have also checked in
examples that the \cond s are preserved under the \pltp y as well.

\section{Invariance of the constraints for the \bc s under the \pltp y}\label{sect5}
{ As we have noted in Sec. \ref{Boundary}, it is not a priori clear what kind of constraints should be
imposed on the gluing operator $\gluingoper$  so that on one hand it properly defines the \bc s as 
D--branes and on the other hand these constraints are preserved under the \pltp y. The examples in the
previous Section indicate that we may have managed to establish the right set of constraints, namely
(\ref{allcon3})--(\ref{allcon6}). We have shown in \cite{ahs:jmp08} that (\ref{allcon3}) is preserved under
\pltp y. It remains to be shown that the others are invariant under the \pl{} \tfn s as well.}

\subsection{An alternative formulation of the consistency \cond s on the gluing operator}\label{alterform} As it is difficult to find
the \pl \tfn{} of the projector ${\cal N}$ it is convenient to reformulate the conditions
(\ref{allcon1}--\ref{allcon6}) without {its explicit use, i.e.,}  using the gluing operator $\gluingoper$ only. This
will also prove that the conditions (\ref{allcon1}--\ref{allcon6}) do not depend on the non--unique choice of the
projector ${\cal N}$ and that we don't have to impose the condition (\ref{allcon5}).

For this purpose we recall Eq. (\ref{Nprop})
$$ \Ran \,{\cal N} = \Ran \,(\gluingoper + id) $$
which means that any condition of the form
$$ {\cal A} \circ {\cal N} =0, \ \ {\rm i.e.,} \ \ N\cdot A=0 $$
can be equivalently written as
$$ {\cal A} \circ (\gluingoper + id) =0, \ \ {\rm i.e.,} \ \ (\gluingmgen + \unit)\cdot A=0. $$
Consequently, the condition (\ref{allcon4}) can be equivalently written as
\begin{equation}\label{allcon4eq}
 (\gluingmgen + \unit) \cdot \left( ({\cal F}+\Delta ) - ({\cal F}+\Delta )^t \cdot \gluingmgen^t \right)  =
0.
\end{equation}
Similarly, the condition (\ref{allcon6}), which when expressed in the basis--free form reads
$$ \rd \Delta ({\cal N} (X),{\cal N} (Y),{\cal N} (Z)) =0, \; \forall X,Y,Z \in T_g \brane, $$
can be equivalently written as
\begin{equation}\label{allcon6eq}
 {(\gluingmgen + \unit)_\kappa}^\nu {(\gluingmgen + \unit)_\lambda}^\rho {(\gluingmgen + \unit)_\mu}^\sigma \partial_{[\nu} \Delta_{\rho \sigma]} = 0.
\end{equation}

Besides that, we recall that the condition (\ref{allcon2}) is just a statement that the distribution (of
non--constant dimension)
$$ \Lambda: g\in G \rightarrow \Ran \,(\gluingoper + id)|_g \subseteq T_g G $$
is in involution,
\begin{equation}\label{allcon2eq}
[\Lambda,\Lambda] \subseteq \Lambda
\end{equation}
and consequently by Frobenius Theorem completely integrable. Such a statement is obviously independent of
the particular choice of the projector ${\cal N}$ (although it doesn't have the nice form $0=\ldots $ of Eq.
(\ref{allcon2})).

Finally,  we look for the the 2--form $\Delta$. We notice that by virtue of Eq. (\ref{allcon3}) the matrix
$$(\gluingmgen + \unit) \cdot \left( {\cal F} - {\cal F}^t \cdot \gluingmgen^t \right)$$
is skew--symmetric and consequently has the form
$$(\gluingmgen + \unit) \cdot M \cdot (\gluingmgen + \unit)^t$$
for some antisymmetric matrix $M$ related to ${\cal F},\gluingmgen$ (and, in general, non--unique).
Therefore, the condition (\ref{allcon4eq}) takes the form
\begin{equation}\label{allcon4eqa}
(\gluingmgen + \unit) \cdot (\Delta + M) \cdot (\gluingmgen + \unit)^t=0
\end{equation}
and, when considered as an equation for $\Delta$, has a solution, e.g. $\Delta=-M $.

Moreover, we can show that the condition (\ref{allcon6eq}) doesn't depend on the particular choice of a solution of
the equation (\ref{allcon4eqa}). It suffices to consider
$$ \Upsilon = (\gluingmgen + \unit) \cdot \Delta \cdot (\gluingmgen + \unit)^t = (\gluingmgen + \unit) \cdot \left( {\cal F}^t \cdot \gluingmgen^t- {\cal F}  \right)$$
and compute the expression
$$ \partial_{\vartheta} \Upsilon_{[\mu \nu} {(\gluingmgen+\unit)_{\lambda]}}^{\vartheta}$$
using the two ways of expressing $\Upsilon$. Due to the integrability condition (\ref{allcon2eq}) written in
terms of generators ${(\gluingmgen+\unit)_{\nu}}^\sigma
\partial_\sigma$ of the distribution $\Lambda$ there exist functions ${\gamma_{\mu \nu}}^{\kappa}$ such that
$$ \partial_\vartheta {(\gluingmgen+\unit)_{[\nu}}^\sigma {(\gluingmgen+\unit)_{\mu]}}^\vartheta = {\gamma_{\mu \nu}}^{\kappa} {(\gluingmgen+\unit)_\kappa}^{\sigma}.$$
Using this fact one finds by comparison of different expressions for $\partial_{\vartheta} \Upsilon_{[\mu
\nu} {(\gluingmgen+\unit)_{\lambda]}}^{\vartheta}$ that
$$
 {(\gluingmgen + \unit)_{[\mu}}^\rho {(\gluingmgen + \unit)_\nu}^\sigma {(\gluingmgen + \unit)_{\lambda]}}^\vartheta \partial_{\vartheta} \Delta_{\rho \sigma} =
$$
$$
\partial_\vartheta \left( {\cal F}^t \cdot \gluingmgen^t -{\cal F} \right)_{\rho [\nu} {(\gluingmgen + \unit)_\mu}^\rho {(\gluingmgen + \unit)_{\lambda]}}^\vartheta - \left( {\cal F}^t \cdot \gluingmgen^t -{\cal F} \right)_{\rho [\nu} \frac{\partial}{\partial y^\vartheta} {(\gluingmgen + \unit)_{\mu}}^\rho {(\gluingmgen + \unit)_{\lambda]}}^\vartheta
$$
(note that index $\vartheta$ in $\frac{\partial}{\partial y^\vartheta}\equiv \partial_\vartheta$ is not antisymmetrized, the
antisymmetrization on the right hand side involves $\mu,\nu,\lambda$ only).

To sum up, we have found that an equivalent formulation of the condition (\ref{allcon6eq}) which doesn't depend on the
particular choice of $\Delta$ exists and has the form
\begin{equation}\label{allcon6eqa}
 \partial_\vartheta \left( {\cal F}^t \cdot \gluingmgen^t -{\cal F} \right)_{\rho [\nu} {(\gluingmgen + \unit)_\mu}^\rho {(\gluingmgen + \unit)_{\lambda]}}^\vartheta - \left( {\cal F}^t \cdot \gluingmgen^t -{\cal F} \right)_{\rho [\nu} \frac{\partial}{\partial y^\vartheta} {(\gluingmgen + \unit)_{\mu}}^\rho {(\gluingmgen + \unit)_{\lambda]}}^\vartheta =0.
\end{equation}
We mention that although the functions ${\gamma_{\mu \nu}}^{\kappa}$ do not appear in the final expression
their existence was important in intermediate steps, i.e. the conditions (\ref{allcon6eq}) and
(\ref{allcon6eqa}) are equivalent only if the integrability condition (\ref{allcon2eq}) holds.

{Watchful reader may notice that we have not imposed the condition (\ref{allcon5}) yet. This condition
restricts the field strength $\Delta$ only to the physically relevant degrees of freedom and it is
reasonable to apply it from this viewpoint. On the other hand, it requires the knowledge of the explicit
form of the projector ${\cal N}$ which we want to avoid. Under the assumption} that the conditions
(\ref{allcon2eq}),(\ref{allcon6eqa}) hold we take {any projector ${\cal N}$ and any} $\Delta$ satisfying
(\ref{allcon4eq}) and construct
$$ \tilde \Delta = N\cdot\Delta\cdot N^t $$
which also satisfies the conditions (\ref{allcon4eq}),(\ref{allcon6eq}) and in addition it satisfies {the
condition (\ref{allcon5}). The influence} of $\Delta$ and $\tilde \Delta$ on the motion of strings, i.e.
extrema of the action (\ref{charged_action}), is exactly the same. Therefore we may consider $\Delta$ and
$\tilde \Delta$ physically equivalent and forget the condition (\ref{allcon5}) altogether.

{In summary} we may write all conditions defining a consistent gluing operator $\gluingoper$ as
\begin{eqnarray}
\gluingmgen \cdot {\cal G}\cdot  \gluingmgen^t & = & {\cal G}, \label{allcon1bis} \\ \label{allcon2bis}
{[\Lambda,\Lambda]} \subseteq \Lambda, \qquad \Lambda(g)&=&\Ran (\gluingoper+id)|_g,\\ \nonumber
\partial_\vartheta \left( {\cal F}^t \cdot \gluingmgen^t -{\cal F} \right)_{\rho [\nu} {(\gluingmgen + \unit)_\mu}^\rho {(\gluingmgen + \unit)_{\lambda]}}^\vartheta - & & \\
  - \left( {\cal F}^t \cdot \gluingmgen^t -{\cal F} \right)_{\rho [\nu} \frac{\partial}{\partial y^\vartheta}
 {(\gluingmgen + \unit)_{\mu}}^\rho {(\gluingmgen + \unit)_{\lambda]}}^\vartheta & = & 0 .\label{allcon3bis}
\end{eqnarray}
Given such an operator $\gluingoper$ we can find the field strength $\Delta$ (using Eq. (\ref{allcon4eq}))
and the projector ${\cal N}$ such that conditions (\ref{allcon3})--(\ref{allcon6}) hold. Both ${\cal N}$ and
$\Delta$ are in general non--unique but lead to the same dynamics of the strings on the classical level, i.e.,
the extrema of the action (\ref{charged_action}).

\subsection{Lift of D-branes to the \dd} We can define the lift of a D-brane $\brane\subset G$ given by (\ref{cphi}) to
the \dd{} as an integral manifold of the distribution generated by\begin{equation}\label{deflift}
    \partial_\tau l\bndry=\partial_- l\bndry+\partial_+ l\bndry.
\end{equation}From $l=g\tilde h$, (\ref{emrpmtilh}) and (\ref{adgt}) we get
\begin{eqnarray}
  \nonumber\partial_\tau l\,l\-1 &=& \left(\rmg+\rpg\right)\cdot T+(\rmtilh+\rptilh)\cdot Ad(g)(\tilde T) \\
  \nonumber&=& (\rmg+\rpg)\cdot T+(\rmg\cdot F(g)-\rpg\cdot F^t(g))(a^{-t}(g)\cdot b^t(g)\cdot T+\tilde T)
\end{eqnarray}
On the boundary we get from (\ref{origbc}), (\ref{Fg}) and (\ref{constH}) \begin{eqnarray}\nonumber
  \partial_\tau l\,l\-1\bndry&=&\rpg\bndry\cdot F^t(g)\cdot\\
  \nonumber&& [(F^{-t}(g)+C\cdot F^{-t}(g))\cdot T+(C-\unit)\cdot(a^{-t}(g)\cdot b^t(g)\cdot T+\tilde T)]\\
  \label{dtllinv}
    &=&\rpg\bndry\cdot F^t(g)\cdot [(E_0^{-t}+C\cdot E_0\-1)\cdot T+(C-\unit)\cdot\tilde  T]
\end{eqnarray}
As $\rpg\bndry$ is arbitrary and $F(g)$ is invertible we see that the vectors tangent to the lifted D-branes
pulled to the unit of the \dd{} form the vector subspace
$\isospace$ of $\cd$
\begin{equation}\label{isospace}
    \isospace = {\rm span}(A^{ab} T_b+{B^a}_b\tilde T^b),
\end{equation}where the matrices $A$ and $B$ are
\begin{equation}\label{matab}
    A=E_0^{-t}+C\cdot E_0\-1,\ \ B=C-\unit.
\end{equation} This subspace is isotropic because\begin{equation}\label{visotropic}
    \langle (A\cdot T+B\cdot\tilde T)^t,A\cdot T+B\cdot\tilde T \rangle= C\cdot E_0^{-t}\cdot C^t-E_0^{-t}
    +C\cdot E_0^{-1}\cdot C^t-E_0^{-1}=0
\end{equation}due to (\ref{cece}). Moreover one can see that the subspace is maximally isotropic as the block matrix
\begin{equation}
    \left(\matrix{A \cr B}\right)=\left(\matrix{E_0^{-t}+C\cdot E_0\-1 \cr C-\unit}\right)
\end{equation}has the same rank as the block matrix
\begin{equation}
    \left(\matrix{E_0^{-t}+E_0\-1 \cr C-\unit}\right),
\end{equation} whose rank is dim $\cg$, because $E_0^{-t}+E_0\-1=E_0\-1\cdot (E_0+E_0^t)\cdot E_0^{-t}
=E_0\-1\cdot {\cal G}(e)\cdot E_0^{-t}$ is an invertible matrix.

The space $\isospace$ is invariant under the \pl{} \tfn{} by construction, nevertheless, one may check it
directly from the \tfn{} properties of $T,\tilde T, E_0$ and $C$. We shall show that the condition
(\ref{allcon3bis}) for admissible gluing matrix $\gluingmgen$ is equivalent to a statement that the
isotropic subspace $\isospace$ is also a subalgebra.

First of all we shall rewrite the matrices occurring in (\ref{allcon3bis}) in terms of the matrices
(\ref{matab}) defining the space $\isospace$.
\begin{equation}\label{ap1}
    \gluingmgen+\unit=\cf^t\cdot (\ca+\cb\cdot\Pic),\ \ \cf^t\cdot\gluingmgen^t-\cf=\cb^t\cdot\cf,
\end{equation}
where
\begin{equation}\label{defacbc}
    \ca=e^{-t}(g)\cdot A\cdot e\-1(g),\ \ \cb=e^{-t}(g)\cdot B\cdot e^t(g),\ \ \Pic=e^{-t}(g)\cdot \Pi(g)\cdot
    e\-1(g).
\end{equation}
The condition (\ref{allcon3bis}) then acquires the form\begin{equation}\label{ap2}
    {\left[\cf^t\cdot\left( \ca+\cb\cdot\Pic \right)\right]_{[\lambda}}^\rho\left[\cf^t\cdot\left( \ca\cdot\dr\cb^t-\dr\ca\cdot\cb^t
    -\cb\cdot\dr\Pic\cdot\cb^t  \right)\cdot \cf\right]_{\mu\nu]}=0.
    \end{equation} (Many terms occurring during derivation of this expression cancel by
    total antisymmetrization in $\lambda,\mu,\nu$.)
    Using (\ref{metorze}) and the fact that both $e(g)$ and $F(g)$ are invertible
    we can simplify the above equation to
\begin{equation}\label{ap3}
    [(A+B\cdot\Pi(g))\cdot e\-1(g)]^{[a\rho}\left( \left[ 2\left(A+B\cdot\Pi(g)\right)\cdot e\-1(g)\cdot\dr
    e(g)-B\cdot\dr\Pi(g)\right]\cdot B^t\right)^{bc]}=0.
\end{equation}(The antisymmetrization involves only the indices $a,b,c$.)
For the derivatives of $e$ we can use Maurer-Cartan equations, and derivatives of $\Pi(g)$ are
\begin{equation}
\label{drpi}
\dr \Pi^{ik}=
 - {(a\-1)_j}^{i}\,{{\widetilde f}^{jm}}{}_{n}\,{\el_\rho}{}^n {(a\-1)_m}^{k},
\end{equation}
where ${\el_\mu}{}^n$ are components of the left--invariant form $\el(g)=e(g) \cdot a(g)$.
All that gives \begin{equation}\label{ap4}
     (A+B\cdot\Pi(g))^{[ai}\left[{f_{ij}}^k(A+B\cdot\Pi(g))^{bj}{B^{c]}}_{k}
    +{a_i}^r(g){\tilde f^{jk}}_{r}{(B\cdot a^{-t}(g))^b}_j{(B\cdot a^{-t}(g))^{c]}}_{k}\right]=0.
\end{equation}(where we again antisymmetrize in $a,b,c$ only). We define a mixed product on the \dd{}
\begin{equation}\label{mixp}
    \langle \langle\ X,Y,Z\ \rangle\rangle:=\langle\ [X,Y],Z\ \rangle.
\end{equation} It is totally antisymmetric and Ad--invariant. In terms of this mixed product we can write the above
condition as \begin{equation}\label{ap7}
    \langle\langle\ (A\cdot T+B\cdot\Pi(g)\cdot T)^{[a}\,,
    \,(A\cdot T-B\cdot\Pi(g)\cdot T+{B}\cdot\tilde T)^b\,,\,({B}\cdot\tilde T)^{c]}\ \rangle\rangle=0.
\end{equation} The antisymmetry of the mixed product and antisymmetrization in indices $a,b,c$ imply\begin{equation}\label{aas}
     \langle \langle\ X^{[a},Y^b,Z^{c]}\ \rangle\rangle=\langle \langle\ X^{[a},Z^b,Y^{c]}\ \rangle\rangle
     =\langle \langle\ Z^{[a},X^b,Y^{c]}\ \rangle\rangle
\end{equation} that allows to rewrite the
left--hand side of (\ref{ap7}) as
\begin{eqnarray}
\nonumber &\langle\langle\ (A\cdot T)^{[a}\,,
    \,(A\cdot T)^b\,,\,({B}\cdot\tilde T)^{c]}\ \rangle\rangle+  \langle\langle\  (A\cdot T)^{[a}\,,
    \,({B}\cdot\tilde T)^b\,,\,({B}\cdot\tilde T)^{c]}\ \rangle\rangle &\\ \nonumber &-\langle\langle (B\cdot\Pi(g)\cdot T)^{[a}\,,
    \,(B\cdot\Pi(g)\cdot T)^b\,,\,({B}\cdot\tilde T)^{c]}\rangle\rangle+  \langle\langle (B\cdot\Pi(g)\cdot T)^{[a}\,,
    \,({B}\cdot\tilde T)^b\,,\,({B}\cdot\tilde T)^{c]}\rangle\rangle.&
\end{eqnarray}
The last two terms drop out by isotropy of the subalgebra $\tcg$ because they are equal to
\begin{eqnarray}
  -\frac{1}{3}\langle\langle (B\cdot\Pi(g)\cdot T-{B}\cdot\tilde T)^{[a}\,,
    \,(B\cdot\Pi(g)\cdot T-{B}\cdot\tilde T)^b\,,\,(B\cdot\Pi(g)\cdot T-{B}\cdot\tilde T)^{c]}\rangle\rangle  &&\nonumber\\
  \nonumber= \frac{1}{3}\langle\langle\ ({B}\cdot a^{-t}(g)\cdot\tilde T)^{[a}\,,
    \,({B}\cdot a^{-t}(g)\cdot\tilde T)^b\,,\,({B}\cdot a^{-t}(g)\cdot\tilde T)^{c]}\ \rangle\rangle=0.
\end{eqnarray}
The first two terms give
\begin{equation}\label{ap8}
   \frac{1}{3}  \langle\langle\ (A\cdot T+B\cdot\tilde T)^{[a}\,,
    \,(A\cdot T+B\cdot\tilde T)^b\,,\,(A\cdot T+B\cdot\tilde T)^{c]}\ \rangle\rangle=0
\end{equation}
and we can drop the antisymmetrization because of antisymmetry of (\ref{mixp}). Then Eq. (\ref{ap8}) becomes
exactly the statement that the maximal isotropic subspace $\isospace$ is a subalgebra of the \dd, i.e., that
for any $v_1,v_2,v_3\in\isospace$ we have
$$ \langle \langle\ v_1,v_2,v_3\ \rangle\rangle=0.$$

To sum up, we conclude that the condition (\ref{allcon3bis}) is in the case of Poisson--Lie dualizable
models equivalent to the statement that the maximally isotropic subspace $\isospace$ is a subalgebra.
Therefore, the condition (\ref{allcon3bis}) is \pl{} invariant.

We also see that the lifts of D--branes into the \dd{} $D$ acquire the form of cosets $\ld l$ where $\ld$ is
the Lie subgroup of $D$ with Lie algebra $\isospace$ and $l \in D$. {This demonstrates that the gluing
matrix formalism naturally leads to D--branes in \dd{} as devised by C.~Klim\v c\'{\i}k and P. \v Severa in
\cite{klse:pltdosdb}. Obviously,} the D--brane in \dd{} $\ld l$ is an embedded submanifold of $D$ whenever
the condition (\ref{allcon3bis}) is satisfied, irrespective of the condition (\ref{allcon2bis}). That leads
us to a natural hypothesis that in our case of dualizable gluing operators the distribution $\Lambda: g\in G
\rightarrow \Ran \,(\gluingoper + id)|_g$ is integrable by  virtue of the condition (\ref{allcon3bis})
alone.

In order to show that the distribution $\Lambda$ is integrable we define a coset projection map
$$\pi: \ D \rightarrow G: \ l=g\tilde h \mapsto g. $$
The D--brane in $G$ passing through $g_0$ is then obtained as $\pi (\ld g_0 \tilde h_0)$ for some $\tilde
h_0 \in \tilde G$ provided that it is well--defined. That it is indeed so can be seen from the fact that for any $l,l'
\in D$ such that $\pi(l)=\pi(l')$ we obviously have
\begin{equation}\label{pirl}
\pi \circ R_l = \pi \circ R_{l'}
\end{equation}
and consequently for any $\ld l_1$, $\ld l_2$ such that $\pi(\ld l_1) \cap \pi(\ld l_2)\neq \emptyset$ we
find that intersecting D--branes in $G$ coincide, i.e. $\pi(\ld l_1) = \pi(\ld l_2)$, and are
submanifolds. Consequently, $\{\pi(\ld l)| l \in D \}$ form a foliation (of non--constant dimension) of the
group $G$ and the distribution $\Lambda$ consisting of tangent spaces to this foliation is by definition
integrable.

For a more explicit derivation it is sufficient to consider a basis of right--invariant vector fields on $D$
extended from a basis $(e_k(e))$ of $\isospace$ by
$$e_k(l)= (R_l)_* e_k(e)$$
and project them by $\pi_*$
$$ E_k(g) = \pi_* e_k (g \tilde h).$$
Such $E_k$ are well--defined vector fields on $G$, i.e. don't depend on the choice of $\tilde h$, due to
Eq. (\ref{pirl}), and define the distribution $\Lambda|_g={\rm span} \{ E_k(g) \}$ by construction of the
lift. Because $e_k$ close under the commutator, also $E_k$ do so due to $\pi_*( [e_j,e_k]) =
[\pi_*(e_j),\pi_*(e_k)]$ and consequently the distribution $\Lambda$ is integrable.

A further question arises concerning the generality of our description, i.e. {whether any D--brane
configuration described in the language of \cite{klse:pltdosdb} can be expressed in terms of gluing
matrices}. Let us suppose that we are given an arbitrary maximally isotropic subalgebra $\isospace$ of the
\dd{} algebra $\cd$, i.e.
$$ \isospace = {\rm span}\{K^{ab} T_b+{L^a}_b\tilde T^b \} $$
where $K,L$ are arbitrary matrices such that
$$ K\cdot L^t+L\cdot K^t =0 $$
and $ \rank \left( K , L \right) = \dim G. $
Does a matrix $C$ exist such that there is an equivalent description
$$ {\isospace} = {\rm span} \{ A^{ab} T_b+{B^a}_b\tilde T^b \}$$
where
$$   A=E_0^{-t}+C\cdot E_0\-1,\ \ B=C-\unit \ ? $$
The answer is positive provided $L-K \cdot E_0^{-1}$ is regular (invertible) matrix. Indeed, we are
looking for an invertible matrix $S$ such that $S\cdot L=A,\ S\cdot K=B$. We find
$$S=\left(E_0^{-t} + E_0^{-t} \right)\cdot \left(L-K\cdot E_0^{-1} \right)^{-1},$$
and
$$ C = \left( E_0^{-t} + E_0^{-t} \right)\cdot \left(L-K\cdot E_0^{-1} \right)^{-1}\cdot K +\unit. $$
Such matrix $C$ satisfies the condition (\ref{cece}). The singular case when $C$ doesn't exist and we
cannot use the description based on gluing matrices occurs if and only if there is
$v\in\isospace,\ v\neq 0$ such that $\langle v,{\cal E}^{-}\rangle =0$, i.e.,
\begin{equation}\label{singC}
v\in \isospace \cap {\cal E}^{+}  \neq 0.
\end{equation}
This is rather exceptional since both $ \isospace$ and $\cal E^{+} $ are $(\dim G)$--dimensional
subspaces in $(2\dim G)$--dimensional vector space $\cd$.

\section{Conclusions}
We have revisited the bosonic version of conditions (\ref{allcond1})--(\ref{allcond4}) formulated in
\cite{ALZ2} for the gluing matrices defining boundary conditions for open strings. We have investigated them
from the point of view of their invariance under the \pl{} \tfn s defined by the formulas
(\ref{Fghat}),(\ref{E0hat}) and (\ref{gluingmh}).

We have seen that in order to keep the \cond s invariant under the \pl{} \tfn s, it is necessary to
introduce the electromagnetic field $\Delta$ on the D-branes where the boundary conditions are imposed as in
\cite{klse:pltdosdb}. Besides that we have relaxed the \cond{} (\ref{multip1}) for the so--called Dirichlet projector ${\cal Q}$
that projects onto the space normal to the D--brane as it is not invariant under the \pl \tfn s. We suggest
that the proper set of constraints for the gluing matrices is (\ref{allcon3})--(\ref{allcon6}). The
invariance of these constraints under the \pl{} \tfn s was firstly checked in many examples; some of them were
presented in Sec. \ref{3dm}. The invariance was proved in Sec. (\ref{sect5}). 

Of course, one may imagine also other possible generalizations of the conditions (\ref{allcond1})--(\ref{allcond4}). One possible approach (in supersymmetric setting) appeared in \cite{SSW} where the condition (\ref{allcond3}) was not strictly enforced whereas the splitting into Dirichlet and Neumann directions due to (\ref{allcond1})--(\ref{multip1}) was retained (together with a stringent restriction $R^2=1$). However, that paper dealt with Abelian T--duality only. In the context of Poisson--Lie T--duality it seems that the condition (\ref{allcond3}), i.e. (\ref{allcon3}), has a natural geometric interpretation, namely the isotropy of lifted D--branes (\ref{visotropic}), and it was essential in most of our derivations. That's why we consider it indispensable in our setting. The condition (\ref{allcon2}) is an integrability statement, needed for interpretation of D--branes as submanifolds. The conditions (\ref{allcon4}),(\ref{allcon6}) are equivalent to the vanishing of the boundary term in the variation of action (\ref{SFg}) and as such are also necessary (as long as one keeps the action in the form (\ref{SFg})). The condition (\ref{allcon5}) restricts the field strength $\Delta$ to a specific choice from a physically equivalent set -- the physics is not at all influenced by it but it is useful for the uniqueness of $\Delta$. To sum up we believe that all the conditions (\ref{allcon3})--(\ref{allcon6}) should be imposed in Poisson--Lie T--duality context.

To prove the \pl{} invariance of the constraints (\ref{allcon3})--(\ref{allcon6}) it was necessary to
reformulate them to the form (\ref{allcon1bis})--(\ref{allcon3bis}) that does not contain the (non--unique)
projector ${\cal N}$. In the end it turned out that the constraints for the gluing matrices
\begin{equation}\label{constHbis}
    \gluingm(g)=F^t(g)\cdot C \cdot F\-1(g),
\end{equation}where $C$ is a constant matrix which satisfies
\begin{equation}\label{cecebis}
    C\cdot(E_0^{-1}+E_0^{-t})\cdot C^t=(E_0^{-1}+E_0^{-t}),
\end{equation} are equivalent to the condition that the subspace 
\begin{equation}\label{isospacebis}
    \isospace = {\rm span}\left(\left(E_0^{-t}+C\cdot E_0\-1\right) \cdot T+{\left(C-\unit\right)} \cdot \tilde T\right)
\end{equation}is a maximally isotropic subalgebra. This statement is clearly invariant under the \pl{} \tfn s because the choice of $\isospace$ is independent of the decomposition of the Lie algebra of the \dd{} into the sum of the isotropic subalgebras
(Manin triple).

On the other hand, if $\isospace$ is a maximally isotropic subalgebra and
$$ \isospace \cap {\cal E}^{+}  = 0$$
then there is a unique matrix $C$ such that $\isospace$ can be written in the form (\ref{isospacebis}) and the
condition (\ref{cecebis}) is satisfied. The gluing matrix (\ref{constHbis}) then satisfies the consistency
\cond s (\ref{allcon1bis})--(\ref{allcon3bis}) or equivalently (\ref{allcon3})--(\ref{allcon6}) where the suitable field strength $\Delta$ is found as a solution of
$$ (\gluingmgen + \unit) \cdot  \Delta \cdot (\gluingmgen + \unit)^t  = (\gluingmgen + \unit) \cdot \left( {\cal F}^t \cdot \gluingmgen^t - {\cal F} \right) $$
and the projector ${\cal N}$ is defined by Eq. (\ref{dfiN}).

This means that we have shown that the current version of the formulation of transformable boundary conditions in terms of gluing matrices is equivalent to the description originally discovered by C.~Klim\v c\'{\i}k and P. \v Severa in \cite{klse:pltdosdb}. Both approaches can be considered complementary. In their original formulation the invariance of the description is clear from its geometric formulation in the \dd{} and also some of the geometric properties of the lifted D--branes are immediately obvious. However, it may be quite tedious to work out the explicit form of the boundary conditions in the \sm s on the groups $G,\hat G$. (E.g. in the original paper \cite{klse:pltdosdb} only the \pl T--duals of free boundary conditions were worked out in any detail. More complicated D--branes in WZW models found in this way were given in \cite{kl:dbeatd}.) On the other hand, in our approach these are easy to write down but it required some calculation to show that both the original and transformed boundary conditions satisfy the same consistency requirements (\ref{allcon1bis})--(\ref{allcon3bis}).

Finally, we would like to recall that we have expressed the conditions on gluing matrix in a form independent of the projector ${\cal N}$, i.e. (\ref{allcon1bis})--(\ref{allcon3bis}), and that this derivation does not depend at all on the particular structure of \pl{} transformable models or on the fact that we consider group targets. We believe that this formulation may be of use also in other investigations of the properties of gluing matrices.

\acknowledgments{Some of the ideas that lead into this work originated in joint discussion
with Cecilia Albertsson for which we are thankful.}


\begin{thebibliography}{00}
\bibitem{ahs:jmp08} C. Albertsson, L.~Hlavat\'{y} and L.~\v{S}nobl, \emph{On the Poisson--Lie T--plurality
of boundary conditions}, \jmp{49}{2008}{032301}, [\arXivid{0706.0820}].

\bibitem{klse:dna}
C.~Klim\v{c}\'{\i}k and P.~\v{S}evera, \emph{Dual non--{A}belian duality and the {D}rinfeld double}, \plb{351}{1995}{455}, [\hepth{9502122}].

\bibitem{unge:pltp} R. von Unge, \emph{Poisson--Lie T--plurality}, \jhep{07}{2002}{014}, [\hepth{0205245}].


\bibitem{ALZ1}
C.~Albertsson, U.~Lindstr\"om and M.~Zabzine, \emph{N=1 supersymmetric sigma model with boundaries, I},
\cmp{233}{2003}{403}, [\hepth{0111161}].

\bibitem{ALZ2}
C.~Albertsson, U.~Lindstr\"om and M.~Zabzine, \emph{N=1 supersymmetric sigma model with boundaries, II},
\npb{678}{2004}{295}, [\hepth{0202069}].

\bibitem{ALZ3}
C.~Albertsson, U.~Lindstr\"om and M.~Zabzine, \emph{T-duality for the sigma model with boundaries},
\jhep{12}{2004}{056}, [\hepth{0410217}]. 

\bibitem{ARE} C. Albertsson and R.A. Reid-Edwards, \emph{Worldsheet boundary conditions  in the
Poisson--Lie T--duality}, \jhep{03}{2007}{004},  [hep-th/0606024].

\bibitem{klse:pltdosdb}
C.~Klim\v c\'{\i}k, P. \v Severa, \emph{Poisson--Lie T--duality: Open strings and D-branes},
 \plb{376}{1996}{82}, [\hepth{9512124}].

\bibitem{klse:osdbwm}
C.~Klim\v c\'{\i}k, P. \v Severa, \emph{Open Strings and D-branes in WZNW model}, \npb{488}{1997}{653}, [\hepth{9609112}].

\bibitem{stanciu} S.~Stanciu, \emph{D-branes in group manifolds}, \jhep{01}{2000}{025}, [\hepth{9909163}].

\bibitem{tvu:pltdpi}
E.~Tyurin and R.~von Unge, \emph{Poisson-Lie T-duality: the
  path-integral derivation}, \plb{382}{1996}{233}, [\hepth{9512025}].

\bibitem{snohla:ddoubles}
L.~\v{S}nobl and L.~Hlavat\'y, \emph{Classification of 6-dimensional
  real Drinfel'd doubles}, \ijmpa{17}{2002}{4043}.

\bibitem{kl:dbeatd}
C.~Klim\v c\'{\i}k, \emph{D--branes in the Euclidean $AdS_3$ and T--duality},
\ijtp{46}{2007}{2443}, [\hepth{0211177}].

\bibitem{SSW}
A. Sevrin, W. Staessens and A. Wijns, \emph{The world-sheet description of A and B branes revisited},
\jhep{11}{2007}{061}, [\arXivid{0709.3733}].

\end{thebibliography}
\end{document}